\documentclass{aa}

\usepackage{graphics}

\def\spose#1{\hbox to 0pt{#1\hss}}
\def\gsim{\mathrel{\spose{\lower 3pt\hbox{$\mathchar"218$}}
          \raise 2.0pt\hbox{$\mathchar"13E$}}}
\def\lsim{\mathrel{\spose{\lower 3pt\hbox{$\mathchar"218$}}
          \raise 2.0pt\hbox{$\mathchar"13C$}}}

\begin{document}

\title{Was the ``naked burst'' GRB 050421 really naked?}

\author{R. Hasco\"et, Z. L. Uhm, R. Mochkovitch, F. Daigne\thanks{Institut Universitaire de France}}
\offprints{R. Hasco\"et (hascoet@iap.fr)}
\institute{Institut d'Astrophysique de Paris, UMR 7095  
Universit\'e Pierre et Marie Curie-Paris 6 -- CNRS,
98 bis boulevard Arago, 75014 Paris, France}

\abstract
{A few long gamma-ray bursts such as GRB 050421 show no afterglow
emission beyond the usual initial steep decay phase. It has been suggested that these 
events correspond to ``naked'' bursts that occur in a very low density environment.
We reconsider this possibility in the context of various scenarios for the origin
of the afterglow.}
{In the standard model where the afterglow results from the forward shock as well as
in the alternative model where the afterglow comes from the reverse shock, we aim to 
obtain constraints
on the density of the environment, the microphysics parameters, or
the Lorentz factor of the ejecta, which are imposed by the absence of a detected afterglow.} 
{For the two models we compute the afterglow evolution for different values of
the external density (uniform or wind medium) and various burst parameters. We then
compare our results to the \textit{Swift} data of GRB 050421, which is the best example of a long burst without
afterglow.}
{In the standard model we show that consistency with the data imposes
that the external density does not exceed $10^{-5}$ cm$^{-3}$ or that the 
microphysics parameters are very small with $\epsilon_e\lsim 10^{-2}$ and
$\epsilon_B \lsim 10^{-4}$. If the afterglow is caused by the reverse shock, we find that its contribution
can be strongly reduced if the central source has mainly emitted fast-moving material
(with less than 10 - 30\% of the kinetic energy at $\Gamma<100$) and was located in a dense environment.} 
{The two considered scenarios therefore lead to opposite constraints on the circumburst medium. The 
high-density environment, favored by the reverse shock model, better corresponds to what is expected  
if the burst progenitor was a massive star.   }

\keywords{Gamma rays burst: general; Gamma rays burst: individual: GRB 050421; Shock waves; 
Radiation mechanisms: non thermal}

\authorrunning{Hasc\"oet et al}
\titlerunning{Was the ``naked burst'' GRB 050421 really naked?}
\maketitle

\section{Introduction}
In the pre-{\it Swift} era afterglow observations typically started a few
hours after the trigger, so that the very early evolution immediately following  
the prompt phase remained a ``terra incognita''. The situation 
dramatically changed with {\it Swift} (Gehrels et al. 2004) which is capable to slew in one minute and
point its X-ray and optical telescopes (XRT, Burrows et al. 2005 and UVOT, Roming et al. 2005) 
to the source. {\it Swift} has revealed
several unexpected features in the early afterglow of gamma-ray bursts (Nousek et al. 2006; O'Brien et al. 2006; Zhang et
al. 2007). The 
prompt phase ends with a steep decay of the X-ray flux,     
$F_{\rm X}\propto t^{-\alpha}$ with $2\lsim \alpha \lsim 5$. The  
afterglow continues with a plateau where the index $\alpha$
lies between 0 and 1. At 0.1 -- 1 day it recovers the more standard value 
$\alpha \sim 1\,-\, 1.5$, which was known before {\it Swift}. Finally, at later times it sometimes
further steepens as a result of a jet break. 
Flares with short rise
and decay times can be superimposed on this global evolution (Chincarini et al. 2007; Falcone et al. 2007).
These different components are not always present.
Flares are observed in about 50\% of the
bursts. The plateau is sometimes absent and the
afterglow then follows a single power-law already from the beginning of the
XRT observations (the most extreme case
being GRB 061007 which maintained a constant slope $\alpha=1.6$ from 100 s, to more 
than 10 days after trigger; Schady et al. 2007).      

GRB 050421 was even more peculiar because it only showed the initial steep
decay phase and a few flares at 100 - 150 s but no plateau and no standard afterglow at later times.  
This behavior had been predicted by Kumar \& Panaitescu (2000) 
for a burst occurring in an extremely low
density environment. In such a ``naked'' burst one only sees the high 
latitude emission once the on-axis prompt emission has stopped.
Radiation from an annulus making an angle $\theta$ with the line of sight arrives with a delay
to the observer and benefits less from the Doppler boost of the relativistic motion.
The predicted flux at a given frequency then decays steeply as $F_{\nu}(t)\propto t^{-\alpha}\nu^{-\beta}$ with
$\alpha=2+\beta$ and $0\lsim\beta\lsim 2$.

In their detailed study of GRB 050421  Godet et al. (2006) found that this event fits well 
with these theoretical predictions and concluded that it was a good naked burst candidate. 
However, the authors did not provide any estimate of the maximum external density that could still 
be compatible with the data. 

A very low density environment has been frequently invoked to explain why 
a fraction of the short burst population has very dim afterglows (see Nakar, 2007, and references therein). 
If short bursts result from
the merging of two compact objects, the kick received when the neutron star or black hole components 
formed in supernova explosions allows the system to reach the low-density outskirts of the host
galaxy  before coalescence occurs.
But GRB 050421 lasted about 10 s and may be 
associated to the long burst population (except 
if it was located at a high redshift, $z>4$; Xiao \& Schaefer, 2011). Long bursts are expected to form 
during the collapse and explosion of rapidly rotating Wolf-Rayet stars (Woosley, 1993). 
The typical environment of the burst should then first
consist of the wind from the star, followed by a wind termination shock
and several shells, successively containing the shocked wind and the remnants of 
previous mass loss episodes (van Marle et al. 2005; Eldridge et al. 2006). 
This may seem to contradict afterglow modeling, which generally favors a
uniform external medium, even for long bursts. One should keep in mind, however,     
that this conclusion relies on several uncertain assumptions such as the constancy of the
microphysics redistribution parameters $\epsilon_e$ and $\epsilon_B$, while the presence 
of a wind is a conspicuous feature in observed Wolf-Rayet stars.

Apart from GRB 050421, at least three other, possibly long bursts (GRB 070531, GRB 080727A and GRB 081016B) 
showed no afterglow
after the steep decay phase (Vetere et al. 2008). GRB 070531 lasted 44 s and had a FRED shape. 
GRB 080727A and GRB 081016BA 
had respective durations $t_{90}=4.9$ and 2.6 s. Because their redshift is not known, it is not clear if they belong
to the short or long burst populations.   

In this work we concentrate on GRB 050421, which has the best data.
Our aim is to perform afterglow calculations to obtain for different scenarios
the limits on the external density that are compatible with the absence of an afterglow.
For a given density we also constrain the microphysics parameters $\epsilon_e$ and $\epsilon_B$ and the distribution of the
Lorentz factor in the ejecta. The paper is organized as follows: we briefly summarize the observational data on GRB 050421 in Sect. 2
and estimate the isotropic kinetic energy released by this burst. We consider 
in Sect. 3 several possible origins for the afterglow. First, the standard case, where it is made by
the forward shock propagating in the external medium, then the alternative model
where it comes from the reverse shock that sweeps back into the ejecta, and finally a few more exotic possibilities. 
Our results are discussed in Sect. 4, which is also the conclusion.

\section{GRB 050421: a burst with no afterglow} 
\subsection{Summary of the observational data}
GRB 050421 belongs to the 10\% faintest bursts of the
{\it Swift} sample. Its fluence in the 15 -- 150 keV energy range 
integrated over $t_{90}=10$ s was $S_{15\,-\,150}= 1.1\pm 0.7 \times 10^{-7}$
erg cm$^{-2}$. The light curve during $t_{90}$ approximately had a FRED shape.
It was followed by a weak tail and at least
two flares at 110 and 154 s. Between 15 and 150 keV the prompt spectrum 
can be fitted by
a single power-law $F_{\nu}\propto \nu^{-0.7}$, which suggests that the peak energy
$E_{\rm p}$ was higher than 150 keV (Godet et al. 2006). The XRT was able to follow
the burst from
about 100 s to 1000 s after trigger. Later, in an interval running 
from 5000 to $5\,10^5$ s, the source was not detected,
leading to an upper limit 
$F_{0.3\,-\,10\,{\rm keV}}< 8\,10^{-14}$ erg cm$^{-2}$ s$^{-1}$. 
\footnote{This value was obtained using the upper limit in count rate from the XRT 
repository (Evans et al. 2007) and the count-to-flux conversion factor used in the 
Burst Analyser (Evans et al. 2010).} 
Any long-lasting 
afterglow component, if present, should therefore be very dim,
lying after a few hours about five orders of magnitude below
the flux recorded at 100 s.   

Between 100 and 1000 s the flux exhibited a power law decline
of index $3.1\pm 0.1$ together with a hard-to-soft evolution, 
indicating that the peak energy of the spectrum was probably crossing the 
XRT band during the observations. This strongly suggests 
that what was observed was the high-latitude emission of the last shocked shells
in the ejecta of GRB 050421 (Godet et al. 2006).

\subsection{Constraining the isotropic kinetic energy of GRB 050421}

\begin{table}
\begin{tabular}[t]{|c|c|c|c|c|c|}
\hline 
$z$ & 0.01 & 0.5 & 1 & 2 & 5
\\
\hline
${\cal E}_{\gamma}^{\rm iso}$ ($10^{52}$ erg) & $9.7\,10^{-6}$ & $2.9\,10^{-2}$ & 0.12 & 0.45 & 2.0 \\
\hline
${\cal E}_{\rm K}^{\rm iso}$ ($10^{52}$ erg) & $3.2\,10^{-4}$ & 0.96 & 4.0 & 15 & 68 \\
\hline
\end{tabular}
\caption{Isotropic gamma-ray and kinetic energies of GRB 050421 for different
redshifts and $H_0=70$ km.s$^{-1}$.Mpc$^{-1}$, $\Omega_{\rm M}=0.27$ and $\Omega_{\Lambda}=0.73$. 
The kinetic energy is given for a radiative efficiency of 3\%. It
would be about ten times lower with an efficiency increased to 30\%.}
\end{table}

The isotropic kinetic energy of the burst ejecta at the end of the prompt phase 
(after a fraction $f_{\gamma}$ of the initial amount has been converted to gamma-rays)
is a key ingredient for any afterglow
calculation. Unfortunately, the redshift of GRB 050421 is not known and,  
in a first step, we just estimate the total gamma-ray fluence 
$S_{\gamma}$
from the fluence in the 15 -- 150 keV band. We obtain 
$S_{\gamma}=4.5\,10^{-7}$ erg cm$^{-2}$
assuming that the spectrum is a Band function (Band et al. 1993) with $\alpha=-1.7$, 
$\beta=-2.5$ and $E_{\rm p}=350$ keV. From the fluence we then obtain 
the total energy release in gamma rays as a function of redshift
\begin{equation}
{\cal E}_{\gamma}^{\rm iso}={4\pi\,D_{\rm L}(z)^{2}\,S_{\gamma}\over 1+z}\ ,
\end{equation}
where $D_{\rm L}(z)$ is the luminosity distance. 
The kinetic energy ${\cal E}_{\rm K}^{\rm iso}$ can now be estimated from the 
efficiency $f_{\gamma}$ 
\begin{equation}
{\cal E}_{\rm K}^{\rm iso}={1-f_{\gamma}\over f_{\gamma}}\,{\cal E}_{\gamma}^{\rm iso}\ .
\end{equation}
In the case of internal shocks we have $f_{\gamma}\simeq \epsilon_e\times f_{\rm diss}$,
where $f_{\rm diss}$ is the fraction of the kinetic energy dissipated by the
shocks and $\epsilon_e$ the fraction transferred to electrons and
eventually
radiated (assuming fast cooling electrons). 
We take $f_{\rm diss}\sim 0.1$, which is typical for internal shocks (Daigne \& Mochkovitch, 1998).
To ensure a sufficient global efficiency, it is then necessary to have $\epsilon_e\sim 0.1$ - 1.
We adopt $\epsilon_e= 1/3$, which leads to
$f_{\gamma}\sim  3\%$. We also consider the possibility that the 
prompt emission may result from a more efficient process such as Comptonization at the
photosphere (Rees \& M\'esz\'aros, 2005; Lazzati, Morsony \& Begelman, 2009; Beloborodov, 2010) 
or magnetic reconnection (Spruit, Daigne \& Drenkhan, 2001; 
Drenkhan \& Spruit, 2002; Giannos \& Spruit, 2006; McKinney \& Uzdensky, 2011), for which we adopt 
a radiative efficiency of 30\%. 
Our results for ${\cal E}_{\gamma}^{\rm iso}$ and ${\cal E}_{\rm K}^{\rm iso}$ are summarized in Table 1 for different redshifts.
They can vary by up to 50\% if the parameters of the
Band function (especially $E_{\rm p}$) are changed. This uncertainty  
remains much smaller than the one resulting from the unknown distance 
and radiative efficiency of the burst.

\section{Explaining the lack of a regular afterglow}
\subsection{The forward shock case}
In the standard model, where the afterglow is made by the forward shock,
the predicted X-ray flux is much above the observational
limit as long as the burst parameters keep ``usual'' values.   
This can be checked using the analytical formulae provided by 
Panaitescu \& Kumar (2000).
The relevant radiative regime corresponds to 
$\nu_{\rm X} > \nu_{m}$ (resp. $\nu_{\rm X} >\nu_{c}$) for fast (resp. slow) cooling, 
where $\nu_{m}$, $\nu_{c}$ and $\nu_{\rm X}$ are the synchrotron, cooling, 
and typical X-ray frequencies respectively.
The expression for the flux density is the same in the two cases and also
for either a uniform external medium or a stellar wind (they differ only
by constant factors on the order of 2 or less).
We have
\begin{equation}
\begin{array}{l}
F_{\rm X}(E) \simeq  \\
 0.3\, D_{28}^{-2}\,{\cal E}_{52}^{(p+2)/4}\,\epsilon_{e,\,-1}^{p-1}\,
\epsilon_{B,\,-2}^{(p-2)/4}E_{\rm keV}^{-(p/2)}\,t^{-[(3p-2)/4]}\ \ {\rm Jy}\ ,\\
\end{array}
\end{equation}
where $\epsilon_e$ (in units of $10^{-1}$) and $\epsilon_B$ 
(in units of $10^{-2}$) are the microphysics redistribution parameters,
${\cal E}_{52}$ is the isotropic kinetic energy in units of $10^{52}$ erg,
$D_{28}$ the luminosity distance 
in units of $10^{28}$ cm, $E_{\rm keV}$ the photon
energy in keV and $t$ the time in seconds (both $E_{\rm keV}$ and $t$ are given 
here in the burst rest frame). 
This relation holds at times longer than the deceleration time. Assuming that
this is the case for $t_{\rm obs}>1000$ s, the data require  
$F_{10\,{\rm keV}}$ to be on the order of $2\,10^{-9}$ Jy at 1000 s and smaller 
than $2\,10^{-11}$ Jy at $5\,10^{4}$ s (Evans et al. 2007, 2010). 
For a reference case defined by $z=1$, ${\cal E}_{\rm K}=4\,10^{52}$ erg, 
$\epsilon_e=0.1$,
$\epsilon_B=0.01$ and $p=2.5$ and using Eq.(3) with the rest frame values $E_{\rm keV}=20$ and
$t=500$ and $2.5\,10^4$ s, 
we obtain
$F_{10\,{\rm keV}}=1.6\,10^{-6}$ and $7.3\,10^{-9}$
Jy at observed times 1000 and $5\,10^{4}$ s respectively. 
The predicted X-ray afterglow is therefore
much brighter than the observational limits. Changing the assumed 
redshift has little effect on this result because $F_{\rm X}(E)\propto
{\cal E}_{52}^{(p+2)/4}\,D_{28}^{-2}$, which does not vary much with $z$
for $2<p<3$.

\begin{figure}
\centerline{\resizebox{0.5\textwidth}{!}{\includegraphics{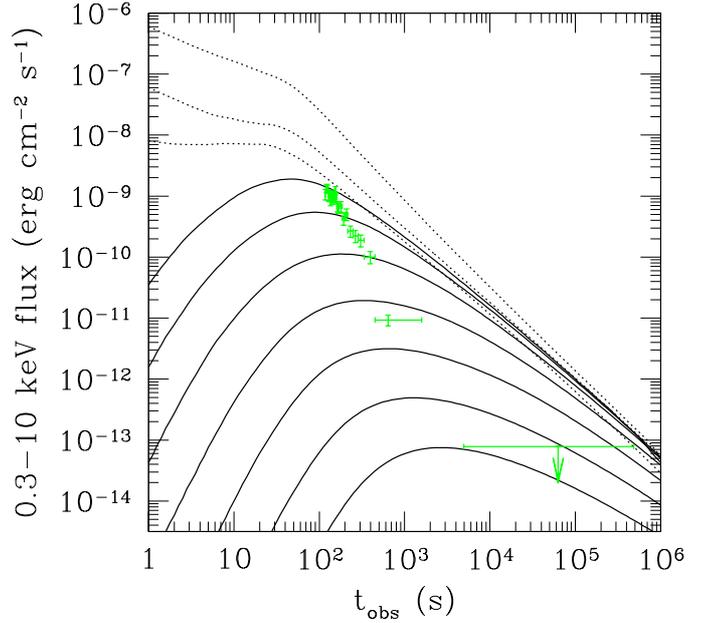}}}
\caption{Afterglow from the forward shock. The assumed redshift and burst energy
are $z=1$ and ${\cal E}_{\rm K}=4\;10^{52}$ erg and the average Lorentz factor of the 
ejecta is ${\bar \Gamma}=150$. The theoretical light curves
in the 0.3 -- 10 keV energy range are presented (from top to bottom)
for $A_*=1$ to $10^{-2}$ (dotted lines) and for $n=1$ to $10^{-6}$ cm$^{-3}$
(full lines). They are compared to the GRB 050421 data from the Burst Analyser (Evans et al. 2010).}
\end{figure}

Another striking consequence of Eq.(3) is that the flux does not depend on
the external density. This remains true as long as 
$\nu_{\rm X}$ is higher than both $\nu_m$ and $\nu_c$. Decreasing the 
density only increases the deceleration time and delays the rise of the
afterglow but does not affect the flux
level in the Blandford-McKee regime. It is only at very low density 
($n<10^{-3}$ cm$^{-3}$)
when $\nu_c$ becomes higher than $\nu_X$ that the radiative regime changes and
the flux begins to depend on density.

We have calculated the evolution of the X-ray flux (in the XRT band
0.3 -- 10 keV) for the  reference case, an average Lorentz factor
in the ejecta ${\bar \Gamma}=150$ \footnote{The choice of ${\bar \Gamma}$ is not critical:
it affects the rise time of the afterglow, but not its evolution in the Blandford-McKee regime.}
and different values of the density: from $n=1$ to $10^{-6}$ cm$^{-3}$ 
(uniform medium)
and $A_*=1$ to $10^{-2}$ (stellar wind). We do not consider lower values
of $A_*$, which would not be realistic for a massive star progenitor. 
Our results are shown in Fig.1. 
It appears that the wind case is clearly excluded while a uniform 
density below $10^{-6}$ cm$^{-3}$ is required, which would likely 
correspond to  
the intergalactic medium (IGM). 
But if GRB 050421 had a massive progenitor 
it should have normally occurred in a 
region of star formation, characterised by a dense environment. With the lower value
of the kinetic energy ${\cal E}_{\rm K}=4\, 10^{51}$ erg 
(for a higher
efficiency of the prompt phase) the maximum allowed density is raised to
about $10^{-5}$ cm$^{-3}$ but still remains very low.

But these conclusions depend on our choice for the 
microphysics parameters. Assuming that $\epsilon_B=\epsilon_e^2$ (which results from
the acceleration process of electrons moving toward current filaments in the shocked material,
Medvedev, 2006),
we find that more standard values of the density ($n>10^{-2}$ cm$^{-3}$ or $A_*>10^{-2}$)
can be made consistent with the data as long as $\epsilon_e<5\,10^{-3}$. Starting from
a lower density, $n=10^{-3}$ cm$^{-3}$, typical of the hot 
interstellar medium and not too far from the transition to the
radiative regime $\nu_m<\nu_X<\nu_c$, the previous limit becomes    
$\epsilon_e<2\,10^{-2}$.
Still with $n=10^{-3}$ cm$^{-3}$ but with the lower kinetic energy ${\cal E}_{\rm K}=4\, 10^{51}$ erg
we finally obtain $\epsilon_e \lsim 4\,10^{-2}$. Except maybe for this final case,
such values of the microphysics parameters are
lower than those usually
inferred from multiwavelength fits of afterglow data (Panaitescu \& Kumar, 
2001a,b; 2002) but it might be possible, for example, that below some 
threshold in density
the transfer of shock-dissipated energy to electrons or/and magnetic
field becomes less efficient, so that $\epsilon_e$ and/or $\epsilon_B$ drop suddenly.
\begin{figure*}
\begin{center}
\begin{tabular}{cc}
\resizebox{0.49\hsize}{!}{\includegraphics{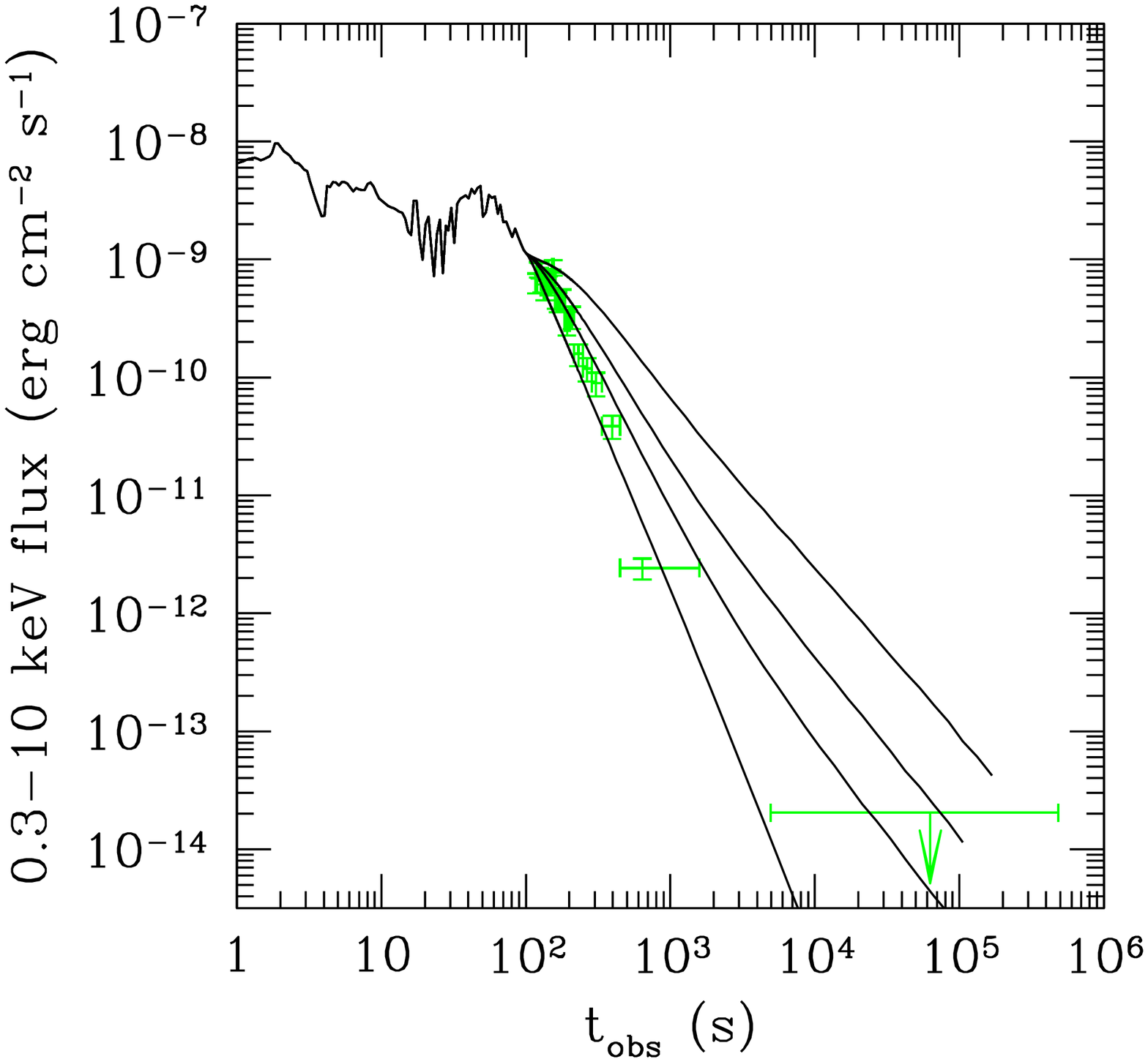}}
\resizebox{0.49\hsize}{!}{\includegraphics{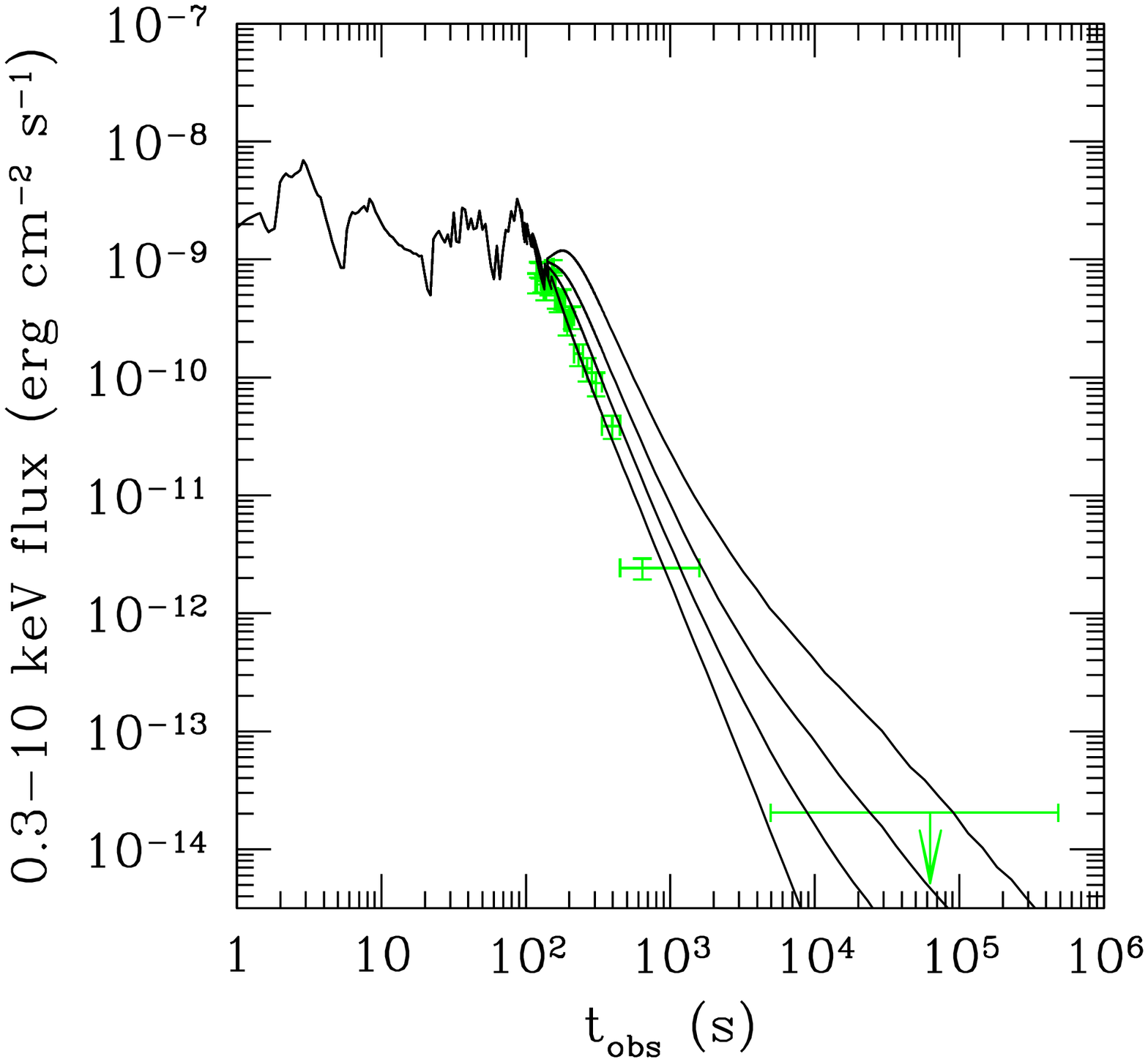}}\\
\end{tabular}
\end{center}
\caption{Afterglow from the reverse shock. Left panel: uniform medium of density $n=1000$ cm$^{-3}$.
Right panel: wind medium with $A_*=1$. The four light curves correspond (from top to bottom) to
${\cal E}_{\rm K}^{\rm slow}/{\cal E}_{\rm K}^{\rm fast}=1$, 0.3, 0.1 and 0,
where ${\cal E}_{\rm K}^{\rm slow}$ (resp. ${\cal E}_{\rm K}^{\rm fast}$) is the kinetic energy in material
with $\Gamma <100$ (resp. $>100$). We assume ${\cal E}_{\rm K}^{\rm fast}=4\;10^{52}$ erg.
} 
\end{figure*}
\subsection{The reverse shock case}
In order to solve some of the problems raised by {\it Swift} observations 
of the early afterglow, Genet et al. (2007) and Uhm \& Beloborodov (2007) have
proposed a non-standard scenario where GRB afterglows 
are made by a long-lived
reverse shock that propagates into the ejecta when it is decelerated 
by the external medium. 
In this scenario it is assumed that the forward shock is present but radiatively 
inefficient (if, for example, the magnetic field is too weak in the external medium) and that
the reverse shock is long-lived because 
the central engine has produced an ejecta with a tail going down 
to very low Lorentz factors (possibly down to $\Gamma \sim 1$). 

The reverse shock model offers an interesting alternative to explain the lack
of an afterglow in objects like GRB 050421, which does not require to have a very low density environment. 
This model assumes that
the central source mainly produced fast-moving material with a limited amount of energy in the tail   
at low $\Gamma$. As it sweeps back into the ejecta, the reverse shock encounters shells with a decreasing energy content
and the observed flux exhibits a steep drop. 

Moreover, because the total energy released by GRB 050421 was 
relatively modest and for a sufficiently high value of the external density,
the reverse shock is relativistic and the emission 
takes place in the fast cooling regime.
This is different from the situation considered by Sari \& Piran (1999) to explain the early
optical flash in GRB 990123, where slow cooling
electrons were responsible for a flux decaying approximately as $t^{-2}$. 
In the present case, a steeper slope can be obtained because 
the light curve is dominated by the high-latitude emission of the last 
shocked shells.  

More precisely, we aim to quantify how much energy we can  
inject into material with a low
Lorentz factor and still remain in agreement with the data. 
To model the source we consider that it has been active 
for $120/(1+z)$ s but that more than 50\% of the total
energy has been released 
during the first $15/(1+z)$ s. This may represent the fact that the
main activity in GRB 050421 had a $t_{90}$ of 10 - 15 s but was followed by a weaker emission
with some flares, lasting for a total of about 100 - 150 s. 
We adopt a distribution of the Lorentz factor that varies between 100 and 400 with a typical
variability timescale of 1 s, which is ended by a tail going 
from $\Gamma=100$ to 2. 

Because we implicitely suppose in this section that the prompt
emission comes from internal shocks, we only consider 
the low-efficiency case for the prompt phase. We then inject
a fixed kinetic energy ${\cal E}_{\rm K}^{\rm fast} = 4\,10^{52}$ erg 
(for $z=1$) into the fast-moving ejecta with $\Gamma>100$ 
and a remaining ${\cal E}_{\rm K}^{\rm slow}$ in the tail ($\Gamma<100$). 
We do not try to fit the details of the prompt light curve 
(which is of poor quality owing to the weakness of the burst) with this distribution but simply to
reproduce the general behavior of the prompt-to-early-afterglow transition.
    
We computed the synchrotron emission from the internal and reverse shocks
as explained in Daigne \& Mochkovitch (1998) and Genet et al. (2007). Because these shocks all take place
in the material ejected by the source and 
are mildly relativistic, we adopted similar
values for the microphysics parameters: 
$\epsilon_e=\epsilon_B=1/3$ and $\zeta$ (fraction of electrons that are
accelerated) $=10^{-2}$, which were also used in the works cited above. 
They ensure a reasonable efficiency in the transfer of dissipated energy to electrons 
and allow the emission to take place in the gamma-ray range during the prompt phase.

The resulting flux in the XRT band is shown
in Fig.2 for four 
values of the ratio
${\cal E}_{\rm K}^{\rm slow}/{\cal E}_{\rm K}^{\rm fast} = 0$, 0.1, 0.3 and 1. 
The density in the burst environment is supposed to be high with 
$n=1000$ cm$^{-3}$ (uniform medium) or $A_*=1$ (stellar wind).
The reverse shock is then relativistic and the emission takes place 
in the fast-cooling regime of the shock-accelerated electrons.

It can be seen in Fig.2 that satisfactory solutions can be found
for both a uniform and a wind external medium as long as the fraction of
energy injected into material with Lorentz factors below 100 does 
not exceed about 10 and 30\% in the uniform and wind medium, respectively.
\footnote{The light curves somewhat differ between the two cases
because owing to the strong deceleration of the ejecta the internal and reverse shocks become mixed.
The profile therefore does not only depend on the distribution of Lorentz factor in the outflow, 
but also on the nature of the environment.}

We checked how these results depend on our assumptions
about the burst redshift and density of the environment. 
Increasing the redshift implies a higher injected energy and shorter
intrinsic time scales. Going to values as high as $z=5$ and keeping the
same density ($n=1000$ cm$^{-3}$ or $A_*=1$) for the environment 
slightly delays the deceleration (in observer time), especially in the 
uniform density case. It is then more difficult to fit the data and 
it could become necessary to inject
essentially the whole energy into material with $\Gamma>200$. 
Similarly, reducing the density of the external medium from
$n=1000$ to $1$ cm$^{-3}$ (at a fixed $z=1$) also delays the deceleration 
and leads to the
same problem.

Therefore, GRB 050421 was not a naked burst in the context of the reverse shock scenario.
On the contrary, it occurred in a dense environnement and was peculiar because it 
released a relatively modest amount of energy, mostly in high Lorentz factor material.
\begin{figure}[t!]
\centerline{\resizebox{0.5\textwidth}{!}{\includegraphics{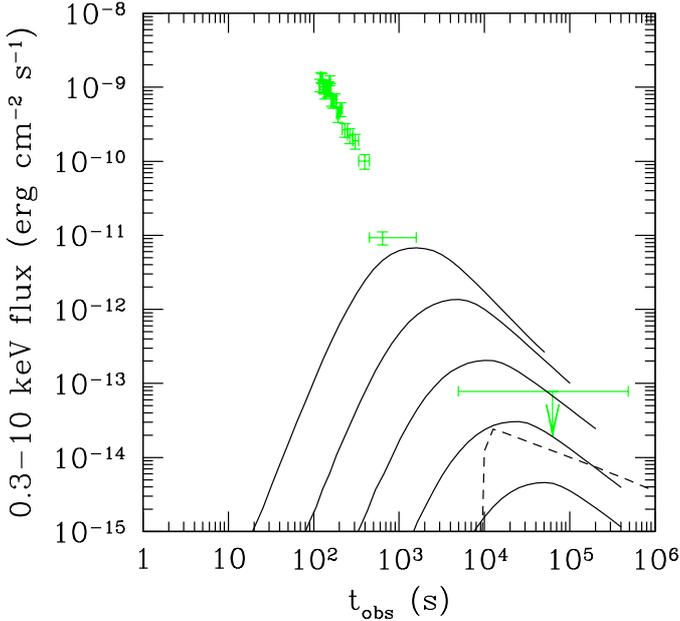}}}
\caption{Afterglow light curves for a sub-luminous burst (full lines) and an interrupted wind
(dashed line). For the sub-luminous burst the injected kinetic energy 
is ${\cal E}_{\rm K}=3.2\,10^{48}$ erg at a redshift $z=0.01$ and the four lines correspond
(from top to bottom) to a density decreasing from 1 to $10^{-4}$ cm$^{-3}$. The 
wind has $A_*=1$ but was interrupted 1000 years before the burst.}
\end{figure}
\subsection{Other possibilities}
\subsubsection{A sub-luminous burst}
It is probable that a large number of sub-luminous bursts
coexists with the classical population of cosmological GRBs. These objects 
are underrepresented in the observed sample because, contrary to the
most powerful events they cannot be detected
at far distances. 

A prototype of these sub-luminous bursts was GRB 980425, which occurred at 34 Mpc
and released an energy ${\cal E}_{\gamma}^{\rm iso} \sim 6\,10^{47}$ erg 
(Galama et al. 1998).
Daigne \& Mochkovitch (2007) argued that GRB 980425 was
intrinsically faint (and not a normal event seen off-axis) and they have  
shown  
that it can be produced in a relativistic outflow with a moderate
Lorentz factor $\Gamma \sim 10$ -- 20. 
This scenario may work for GRB 050421 under the condition that the X-ray afterlow 
becomes dimmer when ${\cal E}_{\rm K}^{\rm iso}$ is decreased at 
constant burst fluence, i.e. if the X-ray flux $F_{\rm X}\propto \left({\cal E}_{\rm K}^{\rm iso}\right)^{x}$ with $x>1$. 
The analytic results of Panaitescu \& Kumar (2000)
show that the most favorable case (with $x=(p+3)/4$) corresponds to 
the radiative regime $\nu_m<\nu_{\rm X}<\nu_c$, in a uniform external 
medium.

Assuming a redshift $z=0.01$ for GRB 050421 we therefore considered an 
outflow with a Lorentz factor between 10 and 15, carrying
a kinetic energy ${\cal E}_{\rm K}^{\rm iso}=3.2\,10^{48}$ erg (see Table 1). 
The resulting afterglow light curves from the forward shock are shown in Fig.3 for $n=1$ to $10^{-4}$ cm$^{-3}$,
$\epsilon_e=0.1$ and $\epsilon_B=0.01$. A hot interstellar medium with
$n\lsim 10^{-3}$ cm$^{-3}$ is almost consistent with the data. Because we have 
$F_{X} \propto \epsilon_e^{3/2}\,\epsilon_B^{7/8}$
(for $p=2.5$), in the considered regime
only a modest reduction of either $\epsilon_e$ or $\epsilon_B$ would be
enough to drive $F_{X}$
below the observational limits. A sub-luminous burst could therefore agree more
easily with the data than a classical GRB 
without implying too low values of the density or 
microphysics  
parameters. But if GRB 050421 was indeed located at $z\sim 0.01$,
one would expect to see a 
candidate host galaxy within one arc minute from the burst and to have detected
an associated supernova. Contrary to GRB 980425, GRB 050421 fails to satisfy
these two criteria.  
 
\subsubsection{An interrupted wind}
We finally consider a more exotic situation where the burst progenitor initially had
a normal stellar wind with $A_*\sim 1$, but we suppose that this wind was 
interrupted 
more than 1000 years before the explosion,
creating a quasi-empty cavity around the star. Because there are no clear justifications
for such a peculiar behavior we only briefly address this case. 
The density in the cavity should not exceed $10^{-5}$ cm$^{-3}$ to ensure
that there will be no afterglow signature before the ejecta hits the
inner end of the wind, located at $R_{\rm w}=3\,v_8\,t_3$ pc where $v_8$ and $t_3$ are
the wind velocity (in units of $10^8$ cm s$^{-1}$) and the time 
during which it 
has been inactive (in units of $10^3$ years). When the ejecta finally reaches $R_{\rm w}$, the wind has expanded to 
the point that the afterglow remains dimmer than the observational limit (see Fig.3). 

\section{Discussion and conclusion}
GRB 050421 was a very peculiar burst with no afterglow after an initial steep
decay phase that went below the XRT detection limit at a few $10^3$ s. This 
behavior corresponds to what is expected for a naked burst occurring in a very low
density environment.
We have reconsidered this interpretation in the context of the standard scenario,
where the afterglow originates from the forward shock, but also within the alternative 
model where it is made by the reverse shock.      

In the first case the density implied for the external medium is indeed very low.
A wind environment with $A_*=0.01\,-\,1$, which would be typical of a Wolf-Rayet progenitor, is clearly
excluded. The limit on the density for a uniform medium somewhat depends on the assumptions 
for the microphysics parameters $\epsilon_e$ and $\epsilon_B$ or the efficiency of the prompt
mechanism, but always remains very low. For standard values, $\epsilon_e=0.1$ and $\epsilon_B=0.01$, 
we obtain $n<10^{-5}$ cm$^{-3}$, lower than any reasonable ISM density and closer to 
a value representative of the IGM.
Conversely, imposing a higher density on the burst environment requires a strong reduction of
the  microphysics parameters, below the values usually found in multiwavelength fits of
afterglow data.  

The fact that only very few long bursts similar to GRB 050421 have been observed
would then be a consequence of the peculiar values required for the burst parameters, i.e. either
an extremely low density environment, or very small $\epsilon_e$ or $\epsilon_B$. These two conditions might 
indeed be related
if below some threshold in density the transfer of shock-dissipated energy to electrons or/and the magnetic
field becomes inefficient. 
Another possibility would be to suppose that GRB 050421 was a short burst and therefore located at $z>4\,-\,5$.
This could more easily account for the low density environment, but the burst should then have released an energy 
${\cal E}_{\gamma}^{\rm iso}$ exceeding $10^{52}$ erg (see Table 1) corresponding to the very upper end of the observed range
for short GRBs (Berger, 2007). 

Still within the scenario where the afterglow comes from the forward shock we briefly considered 
two special cases: in the first one GRB 050421 was a nearby, sub-luminous burst and in the second
it was surrounded, at the moment of the explosion, by a quasi-empty cavity created by a wind that
was interrupted a few thousands years before the burst. Both can be made compatible with the XRT data 
but not with the absence of a host galaxy or supernova imprint in the first case, while the second case
relies on a very ad hoc assumption that lacks clear justification. 

In the alternative reverse shock scenario a long-lasting afterglow emission is produced when 
a tail of material with low Lorentz factor is present in the ejecta emitted by the
central engine. We suggest that in some occasions this tail might be missing, 
which would simply explain the absence of an afterglow in objects like GRB 050421. Moreover, 
to ensure that the observed emission ends with the high-latitude emission of the last
shocked shell, the radiating electrons must be in the fast-cooling regime, which is possible
if the external medium has a high density. The situation  
is then just the reverse from the one found in the standard scenario: a dense burst environment is favored,
as expected if the burst progenitor was a massive star.

\begin{acknowledgements}
We thank the anonymous referee for detailed and helpful comments and 
Paul O'Brien for kindly answering our questions regarding the XRT results. This work is partially supported by the French Space Agency (CNES).
R.H.'s PhD work is funded by a Fondation CFM-JP Aguilar grant and
Z.L.U. is supported by the grant ``Research in Paris 2010/2011'' of the City Hall of Paris.

\end{acknowledgements}


\vfill

\begin{thebibliography}{}

\bibitem{}
Band, D., Matteson, J., Ford, L. et al. 1993, ApJ, 413, 281
\bibitem{}
Beloborodov, A.M. 2010, MNRAS, 407, 1033
\bibitem{}
Berger, E. 2007, ApJ, 670, 1254
\bibitem{}
Burrows, D.N., Hill, J.E., Nousek, J.A. et al. 2005, Space Sci. Rev., 120, 165  
\bibitem{}
Chincarini, G., Moretti, A., Romano, P. et al. 2007, ApJ, 671, 1903
\bibitem{}
Daigne, F., \& Mochkovitch, R. 1998, MNRAS, 296, 275
\bibitem{}
Daigne, F., \& Mochkovitch, R. 2007, A\&A, 465, 1
\bibitem{}
Drenkhahn, G., \& Spruit, H.C. 2002, A\&A, 391, 1141
\bibitem{}
Eldridge, J.J., Genet, F., Daigne, F. et al. 2006, MNRAS, 367, 186
\bibitem{}
Evans, P.A, Beardmore, A.P., Page, K.L. et al. 2007, A\&A, 469, 379 
\bibitem{}
Evans, P.A., Willingale, R., Osborne, J.P. et al. 2010, A\&A, 519, 102
\bibitem{}
Falcone, A.D., Morris, D., Racusin, J. et al. 2007, ApJ, 671, 1921
\bibitem{}
Galama, T.J., Vreeswijk, P.M., van Paradijs, J. et al. 1998, Nature, 395, 670
\bibitem{}
Gehrels, N., Chincarini, G., Giommi, P. et al. 2004, ApJ, 611, 1005
\bibitem{}
Genet, F., Daigne, F., Mochkovitch, R. 2007, MNRAS, 381, 732
\bibitem{}
Giannos, D., \& Spruit, H.C 2006, A\&A, 450, 887
\bibitem{}
Godet, O., Page, K.L., Osborne, J.P. et al. 2006, A\&A, 452, 819
\bibitem{}
Kumar, P., \& Panaitescu, A. 2000, ApJ, 541, L51
\bibitem{}
Lazzati, D., Morsony, B.J., Begelman, M.C. 2009, ApJ, 700, L47
\bibitem{}
McKinney, J.C., \& Uzdensky, D.A. 2011, MNRAS, submitted (arXiv:1011.1904)
\bibitem{}
Medvedev, M.V. 2006, ApJ, 651, L9
\bibitem{}
Nakar, E. 2007, PhR 442, 166 
\bibitem{}
Nousek, J.A., Kouveliotou, C., Grupe, D. et al. 2006, ApJ, 642, 389
\bibitem{}
O'Brien, P.T., Willingale, R., Osborne, J. et al. 2006, ApJ, 647, 1213
\bibitem{}
Panaitescu, A, \& Kumar, P. 2000, ApJ, 543, 66
\bibitem{}
Panaitescu, A, \& Kumar, P. 2001a, ApJ, 554, 667
\bibitem{}
Panaitescu, A, \& Kumar, P. 2001b, ApJ, 560, L49
\bibitem{}
Panaitescu, A, \& Kumar, P. 2002, ApJ, 571, 779
\bibitem{}
Rees, M.J. \& M\'esz\'aros, P. 2005, ApJ, 628, 847
\bibitem{}
Roming, P.W.A., Kennedy, T.E., Mason, K.O. et al. 2005, Space Sci. Rev., 120, 95  
\bibitem{}
Sari, R., \& Piran, T. 1999, ApJ, 517, L109
\bibitem{}
Schady, P., de Pasquale, M., Page, M.J. et al. 2007, MNRAS, 380, 1041
\bibitem{}
Spruit, H.C.; Daigne, F., Drenkhahn, G. 2001, A\&A, 369, 694
\bibitem{}
Uhm, Z.L., \& Beloborodov, A.M. 2007, ApJ, 665, L93
\bibitem{}
van Marle, A.J. Langer, N., Garc{\' i}a-Segura, G. 2005, A\&A, 444, 837
\bibitem{}
Vetere, L., Burrows, D.N., Gehrels, N. et al. 2008,  
GAMMA-RAY BURSTS 2007: Proceedings of the Santa Fe Conference. AIP Conference Proceedings, 
Volume 1000, pp. 191-195
\bibitem{}
Woosley, S.E. 1993, ApJ, 405, 273
\bibitem{}
Xiao, L., \& Schaefer, B.E. 2011, ApJ, 731, 103 
\bibitem{}
Zhang, B., Liang, E., Page, K. L. et al. 2007, ApJ, 655, 989
\end{thebibliography}
\end{document}